# Design and Analysis of Microfluidic Cell Counter using Spice Simulation

Sheikh Muhammad Asher Iqbal[1] and Nauman Zafar Butt[1]

**Abstract:**

Microfluidic cytometers based on coulter principle have recently shown a great potential for point of care biosensors for medical diagnostics. In this study, the design and characterization of coulter based microfluidic cytometer are investigated through electrical circuit simulations considering an equivalent electrical model for the biological cell. We explore the effects related to microelectrode dimensions, microfluidic detection volume, suspension medium, size/morphology of the target cells, and, the impedance of the external readout circuit, on the output response of the sensor. We show that the effect of microelectrodes' surface area and the dielectric properties of the suspension medium should be carefully considered when characterizing the output response of the sensor. In particular, the area of the microelectrodes can have a significant effect on cell's electrical opacity (the ratio of cell impedance at high to low frequency) which is commonly used to distinguish between sub-populations of the target cells (e.g. lymphocytes *vs.* monocytes when counting white blood cells). Moreover, we highlight that the opacity response vs. frequency can significantly vary based upon whether the absolute cell impedance or the differential output impedance is used in its calculation. These insights could provide valuable guidelines for the design and characterization of coulter based microfluidic sensors.

**Keywords:**

Microfluidic Cell Counter, Coulter Principle, Electrical Cell Model.

Nauman Zaffar Butt, nauman.butt@lums.edu.pk | [1]Department of Electrical Engineering, SBA School of Science and Engineering, Lahore University of Management Sciences (LUMS), Lahore, Pakistan.

1. **Introduction:**

The interaction of biological cells with an externally applied electrical field finds many application in medical science [1]. For example, in the process of electroporation, stimulation of cells at a high electrostatic field is used to create pores in the cell membrane which allow for transferring drugs or DNA into the cell [2]. Similarly, in dielectrophoresis, forces exerted on the cell due to an applied AC electric field are used to separate cells having different size or shape. This technique finds various medical applications such as separating cancer cells from healthy cells, platelets from the whole blood, and, live cells from dead cells, *etc.* [3]. A relatively recent application of the electrical interaction of biological cells is in the area of medical diagnostics where microfluidic cell cytometry based on the Coulter principle is used to enumerate biomarkers for a particular disease [4]. Other applications of this technique include purification of liquids by detecting particles of impurities, detecting pollens and drug administering. A typical Coulter based cell cytometer can count or sort cells of various types based on the change in the electrical impedance of a $\sim\mu$-liter volume of an electrolyte as the cells flow across the embedded microelectrodes. In an immunoassay biochip, two identical counters are typically placed, one at the inlet, and, the other at the outlet of a microfluidic immuno-capture chamber. The capture chamber contains antibodies that are specific to the target biomarkers and are usually coated on 3D micro-pillars to enhance the capture efficiency. A differential counting of the biomarker at the entry and the exit of the chamber is used to enumerate the target biomarker. An on-chip electronic circuit for signal processing and data acquisition can potentially be integrated with the microfluidic biochip for the portable biosensor. The main advantages of these electronic biochips are their high precision, label-free detection, portability, and, compatibility with high throughput chip manufacturing process.

One of the prime application of microfluidic cell cytometry is to detect biological cells which leads to diagnosis of fatal diseases like HIV/AIDS and Cancer. Currently, there are about 36.4 million people globally being affected by HIV/AIDS and their number is increasing at a rate of 5000 new infections per day [5]. Around 11 million (30%) of these HIV/AIDS patients did not have access to HIV testing services. A microfluidic point of care device could be extremely valuable for a low cost and accessible solution for such epidemics. In this regard, immunoassays based on Coulter based microfluidic enumeration of biomarkers in a drop of blood sample have recently attracted a lot of attention for point of care (PoC) medical diagnostics [6,7]. A variety of applications have been demonstrated including stratification of sepsis based on the real-time quantification of CD64 and cytokines, and, HIV/AIDS prognostics using CD4 enumeration [7].

Previously, Chen [8] has summarized developments in microfluidic impedance based cytometry in a chronological order. Gawad [9] demonstrated an impedance spectrometry microfluidic cytometer for cell analysis and sorting. The study used 3D finite element simulation to compare different electrode geometries and cell positions. A simplified version of the electrical model for the biological cell was used that was originally introduced by Schoenbach [10] to understand the impedance response of the sensor. Ellappan [11] used Cadence Spectra tool to study the basic response of a biological cell at variable signal frequencies. Jayara [12] optimized the microfluidic channel for the cell growth and described useful approaches to fabricate, optimize, and validate a biocompatible device. This work concluded that an increase in the channel depth could induce surface roughness which could significantly affect the device characteristics. Qiu [13] optimized the microfluidic channel to improve the hydrodynamic dissociation of cells from tissues and organs for cell identification and disease diagnosis. Watkins [6] developed a differential microfluidic cytometry technique for the $CD4^+$ and $CD8^+$ lymphocyte counter for HIV diagnostic applications. Hassan [14] published a detailed protocols for the fabrication of differential microfluidic cytometer biochip. Hassan [15] demonstrated an application of the differential microfluidic cell counter biochip for the complete blood cell count. In another study [7], Hassan used the differential cell counter biochip to characterize CD 64 cells from the whole blood samples for sepsis stratification. More recently, PoC microfluidic immunoassays based on Coulter principle have been applied to enumerate proteins in a drop of whole blood [16].

A number of other microfluidic platforms based on impedance cytometry have been demonstrated for the biodetection and analysis of micro-analytes. For example, Evander [17] discriminated platelets from red blood cells using cellular electrochemical impedance spectroscopy and dielectrophoresis. Haandbaek [18] characterized the subcellular morphology of yeast cell using a novel microfluidic impedance cytometer that is capable of simultaneous analysis at four frequencies between DC and 500 MHz. Haandbaek [19] also detected single bacteria using microfluidic impedance cytometer by incorporating a series resonator circuit that improved the sensor's sensitivity. Song [20] developed a non-invasive, label-free, micropore-based microfluidic impedance flow cytometer for the identification of the differentiation state of the stem cells. Bernabini [21] demonstrated the detection of micro-particles by hydrodynamically focusing the particles in the center of 200μm microfluidic channel. The proposed technique demonstrated a high sensitivity for the microfluidic cytometer while maintaining a large channel dimensions. Liu [22] utilized electrical impedance differential sensing for the detection of sickle cells by combining it with oxygen control onto a single microfluidic chip. Petchakup [23] developed a novel microfluidic strategy by combining the label-free leukocyte sorting with impedance profiling. With this technique, enhanced impedance detection selectivity for diabetes testing was demonstrated. Honrado [24] characterized plasmodium falciparum-infected red blood and differentiated infected and non-infected red blood cells based on their cell membrane properties. Chawla [25] demonstrated an integrated microfluidic platform that can be used to assess the viability of Newly Transformed Schistosomula (NTS) using an impedance-based analysis method. The proposed platform was used for the automated antischistosomal drug screening. Chawla [26] also used microfluidic platform for the long-term culturing and high-resolution imaging of yeast cells. Finally, the microfluidic cytometry has also been used for organ on chip applications, e.g., for cell detection with temporal regulation of the cell microenvironment [27], selective detection of the migratory properties of cancer cells [28], and, for measuring Transepithelial electrical resistance (TEER) [29].

Although a large number of studies have demonstrated the application of microfluidic cytometer for a variety of biomarkers' detection, the design aspects of the cytometer and a systematic analysis of its performance as a function of cellular properties and those of the suspension medium are relatively unexplored in the literature. In particular, a detailed study of the electrical characterization or modeling of the microfluidic cytometer system is desired which could guide the device design and help in the interpretation of experimental data. In this paper, we use electrical circuit simulations to investigate the design insights and guidelines for the optimized structural and circuit parameters of the device. In particular, we study the effects of microelectrode dimensions, medium's impedance and its dielectric properties, the readout circuit's resistance, and, the dimensions of the microfluidic channel on the electrical output of the sensor. Moreover, we explore the electrical response of the sensor as a function of varying the cell properties such as cytoplasm and nuclear impedances, and, membrane capacitance. Finally, the electrical opacity for the cells which can provide useful information about cell membrane properties is discussed in context of the design parameters and the input signal frequency. Some of the unique contributions of our study are: (i) The effect of electrode area on the opacity. Moreover, the significance of the method for defining opacity in the experiments and the related effect of various cell/system parameters on opacity is discussed. (ii) The effect of selecting the external readout resistance on the output signal is quantified, and, (iii) The effect of practical suspension mediums such as PBS, whole blood, DI water etc., on modifying the detection signal is presented.

## 2. Modelling Approach:

The detection of cell biomarker in a microfluidic counter is based on sensing the difference in the impedance inside a detection volume as the cell traverses through it. In the literature, the cell impedance in a suspension medium has been modeled through several approaches [27]. These include the finite element/difference methods [28], equivalent circuit model [29, 30], the boundary element method [31], the transport lattice method [32, 33]. We have used the equivalent circuit model approach in this work. Although this approach is simpler as compared to other approaches, it can be very useful to highlight the qualitative trends important for the design and characterization of the device.

The equivalent circuit model is implemented in LT Spice (LTspice) which is a high performance SPICE (Simulation Program with Integrated Circuit Emphasis) tool widely used for simulating electrical circuits. The output voltage signal from the microelectrode sensor is monitored as a function of broad range of physical design parameters for the applied electrical signal frequency ranging from $100\ Hz$ to $100\ MHz$.

### 2.1. Cell Electrical Model:

Fig.1$a$ shows an illustration of a biological cell that consists of a cell membrane, cytoplasm and nucleus. An equivalent electrical model for the cell [10] is shown in Fig. 1$b$. The cell has been modelled as a uniform conductive medium, i.e., cytoplasm, having a nucleus and an outer cell membrane. The cell membrane has been modelled as a bilayer spherical dielectric having capacitance, $C_m = (\frac{\kappa \varepsilon A}{d})$. Nucleus of the cell is modelled as a double lipid bilayer having capacitance, $C_n$, which is half of that of $C_m$. The cell's cytoplasm is modelled as a resistance, $R_c$. The impedance of the suspension medium (excluding cells) in the microfluidic detection volume is represented by the resistance of solution ($R_{sol}$). A double layer capacitance, $C_{dl}$ appears at the interface between a conductive electrode and an adjacent liquid electrolyte [34]. We take the resistivity of cytoplasm and nuclear plasma to be the same and assume a typical value of 100 ohm-cm [10]. For our reference cell diameter of 15 um, this gives $R_{c1} = 16.7 K\Omega, R_{c2} = 133 K\Omega, R_{c3} = 16.7 K\Omega$, $C_m = 4pF$, and, $C_n = 2pF$. Finally, nuclear resistance is assumed to be equal to the cytoplasm resistance, i.e., $R_n = 133 K\Omega$ (eq. a-d ). Perturbation to these reference values for cell parameters are specified when used. The amplitude (∥) and phase (∠) of the cell impedance, ($Z_{cell}$) as function of frequency ($f$) of the applied voltage are plotted in Fig. $2a$. The detection volume is considered to be equal to the cell volume unless otherwise specified. Since amplitude of a capacitor's impedance scales inversely with frequency, $|Z_{cell}|$ decreases with increasing $f$ until all of the capacitive elements in $Z_{cell}$ becomes negligibly small (i.e., the capacitors behave like a short circuit). For $f > 100 KHz$, the amplitude of $Z_{cell}$ saturates to a minimum value while its phase approaches to zero (a characteristic of a pure resistive circuit).

$$R_{c1} = R_{c3} = \frac{\rho \times L}{A} = \frac{100}{4 \times 15 \times 10^{-4}} \quad (a)$$

$$R_{c2} = \frac{\rho \times L}{A} = \frac{100}{7.5 \times 10^{-4}} \quad (b)$$

$$C_{m1} = C_{m2} = (\frac{9 \times 8.85 \times 10^{-12} \times 4 \times \pi (7.5 \times 10^{-6})^2}{2 \times 7 \times 10^{-9}}) \quad (c)$$

$$C_{n1} = C_{n2} = \frac{C_{m1}}{2} \quad (d)$$

## 2.2. Device Structure:

Fig $1c$ shows an illustration of the microfluidic cell counter. The device contains three coplanar micro electrodes immersed in the microfluidic detection volume. An AC electrical voltage signal of $1V$ amplitude is applied at the middle microelectrode while voltages at the remaining two microelectrodes are detected. The readout circuit comprises of two identical external resistances ($R_{out}$), which form a Wheatstone bridge connection with the microfluidic impedances (Fig.1$d$). The series resistors ($R_{s1}$ and $R_{s2}$) and parallel resistor ($R_p$) in Fig. 1$d$ represents the suspension medium resistances surrounding the cell in the detection volume. For an ideal case, the cell size is equal to the detection volume for which $R_{s1}$ and $R_{s2}$ are assumed zero while $R_p$ is infinite.

## 2.3. Sensing Principle:

A microfluidic counter based on coulter principle senses the cells or polystyrene beads coated with proteins individually as they flow through the microelectrodes in a microfluidic channel having detection volume close to the diameter of a single target cell (or bead). As a cell traverses through the detection volume, it induces an electrical pulse at the microelectrodes due to a difference in the impedance between the biological cell and that of the background electrolyte medium. This electrical pulse is sensed across $R_{out}$. A differential voltage ($\Delta V_{out}$) is sensed at the output of the cell counter by taking the difference between $V_{out1}$ (when detection volume contains a cell) and $V_{out2}$ (when there is no cell in detection volume) measured across $R_{out1}$ and $R_{out2}$ respectively:

$$|\Delta V_{out}| = |V_{out1}| - |V_{out2}| = |(|I_1 - I_2|) \times R_{out}| \quad (e)$$

where $I_1$ and $I_2$ are the AC electrical currents flowing out of left and right microelectrode respectively. Fig $2b$ and $2c$ show an example for the amplitude and phase of $V_{out1}$ and $V_{out2}$ respectively. Since $V_{out2}$ is influenced by $C_{dl}$ and $R_{sol}$ while $V_{out1}$ is influenced by $C_{dl}$ and $Z_{cell}$, their frequency response is different. In particular, $V_{out2}$ for the given design shows saturation at a lower frequency as compared $V_{out1}$. Moreover, at high $f$ ($> 1MHz$), $Z_{cell} > R_{sol}$ for this device, so the saturated amplitude for $V_{out2}$ is greater than that of $V_{out1}$. The resultant characteristics for $\Delta V_{out}$ are shown in Fig 2$d$. From Fig. 2, it is evident that the design of device parameters (which influence $C_{dl}$ and $R_{sol}$) and the cell's size/shape characteristics (which influence its capacitances and resistances) both can affect the sensor's output response. It should be noted that the frequency response of the sensor is significant for the cell analysis because the cell morphology (primarily the membrane capacitance) primarily affects $\Delta V_{out}$ at a specific range of frequency (typically around $1 MHz$). The ratio between the output responses at a higher frequency to that at a lower frequency is therefore often used to extract valuable information about the cell shape and morphology [6, 14, 9]. This aspect is explored in detail in Section 3.

## 3. Results and Discussions:

In this section the simulation and design of the microfluidic cytometer device is discussed in detail. The results and discussion are covered into four parts concerning: (i) Device parameters that include microelectrode dimensions and $R_{out}$, (ii) Suspension medium parameters which includes its dielectric constant and $R_{sol}$, (iii) Microfluidic channel design parameters which includes the dimensions of the detection volume, and (iv) Cell (biomarker) parameters which include electrical parameters related to its size/morphology.

### 3.1. Device Parameters:

Device parameters include optimizing electrode dimensions and external resistance, $R_{out}$.

#### 3.1.1. Optimizing Microelectrode Dimensions:

As the surface area ($A$) of the microelectrodes is directly proportional to $C_{dl}$, it can influence the output voltage response of the sensor. Fig $3a - c$ show that the amplitude and phase of $\Delta V_{out}$ can both indeed be significantly modified with the change in $A$. As the electrode area increases, the amplitude of $\Delta V_{out}$ at lower frequencies is enhanced. Since all the capacitive impedances in the circuit become negligible at high $f$ ($> 1MHz$), $|\Delta V_{out}|$ saturate to a constant amplitude which is determined by $R_{sol}$ and the cell resistances. It can be noted that for the microelectrode area of $200 \times 200\ \mu m^2$, the low frequency peak of $\Delta V_{out}$ is most broad and high. For example, at $f = 10\ kHz$, $\Delta V_{out}$ for $A=200 \times 200\ \mu m^2$ and $A=10 \times 10\ \mu m^2$ is $0.19\ V$ and $0.7\ mV$, respectively, hence a shift of about $0.189\ V$. As $f$ increases to ~$1.5\ MHz$, $|\Delta V_{out}|$ saturates to a maximum of $0.1\ V$ irrespective of $A$. Fig. $3c$ shows $|\Delta V_{out}|$ as function of $A$ for $f = 10kHz, 100kHz$, and, $1MHz$. While the microelectrode area has negligible effect on $|\Delta V_{out}|$ for $f = 1MHz$, it significantly affects $|\Delta V_{out}|$ for $f = 10kHz$. It should be noted that an optimal value of $A$ which can maximize the low frequency response of the sensor is slightly above $A = 10^4 \mu m^2$. A further increase in $A$ above this optimal value does not provide any additional benefit to the low frequency response of the sensor. Very small values (e.g., $< 10^3 \mu m^2$), on the other hand, significantly degrade the sensor's output response for $f < 100kHz$.

#### 3.1.2. Optimizing R<sub>out</sub>:

Fig $3d - f$ shows $\Delta V_{out}$ trends with varying $R_{out}$. It should be noted that the choice of $R_{out}$ not only influences $|\Delta V_{out}|$ at high frequency but could also affect the peak in the low frequency response. As $R_{out}$ is increased above an initial value of $5K\Omega$, $|\Delta V_{out}|$ both at high and low frequencies shows an increased amplitude. Increasing $R_{out}$ above $50K$ however starts to saturate the amplitude of $\Delta V_{out}$ at high frequency but the low frequency response keeps showing an increase in amplitude. Finally, for $R_{out} = 100M\Omega$, the higher frequency amplitude significantly drops while the low frequency amplitude is further increased. From Fig $3d$, it is clear that an optimal choice of $R_{out}$ for an impedance cytometer is significantly dependent on $f$. For example, at $f = 100$ KHz, an optimal $R_{out}$ of $1M\Omega$ provides the largest amplitude for the output signal whereas at a higher frequency, i.e. $1\ MHz$, the optimal $R_{out}$ is $50\ K\Omega$. These results highlight that the choice of $R_{out}$ must be made with consideration of the expected range of operating frequency of the cytometer.

### 3.2. Trends with suspension medium parameters:

The suspension medium parameters include the resistance ($R_{sol}$) and dielectric constant ($\kappa$).

### 3.2.1. Resistance of the suspension medium:

In a practical implementation of the microfluidic cytometer, the cells are counted directly from a lysed blood sample or blood plasma. Other reagents such as phosphate buffer solution ($PBS$) and deionized water ($DI$) etc. can sometimes be incorporated as suspension medium for cells or beads in the laboratory characterization. The value of $R_{sol.}$ can be calculated from the conductivity of the solution medium for a given dimension of the detection volume. Fig 4 shows the effect of $R_{sol}$ on $\Delta V_{out}$. It can been seen that highest $\Delta V_{out}$ is obtained when $PBS\ 10\times$ is the medium due to its high conductivity $12.3\ S/m$ (Lenntech). For a Phosphate Buffered Solution ($PBS\ 1\times$) medium (conductivity of $1.6\ S/m$) and the whole blood sample having conductivity of $1.09\ S/m$ ($PMC$), $\Delta V_{out}$ at frequency $1MHz$ is approximately $80mV$ and $134mV$ respectively. For $DI\ water$ solution, conductivity ($5.5\times\ S/m$ (AmericanBio)), is very low and results in a significant drop in $\Delta V_{out}$.

### 3.2.2. Trends with Dielectric constant:

The dielectric properties of the solution medium affect $C_{dl}$ and hence the electrical response of the cytometer. Fig. $5a$ shows the plot between $\Delta V_{out}$ and the dielectric constants of different electrolyte solutions assuming $A = 15\times 15\mu m^2$.

It can be seen in Fig $5a$ that the effect of increasing $\kappa$ improves the low frequency amplitude of the cytometer output. This trend is similar to what has been shown in Fig. $3a$ for $\Delta V_{out}$ response as a function of $A$. This is expected since both $A$ and $\kappa$ has as similar effect on the $C_{dl}$ at all frequencies of the applied signal. Fig. 5 shows that the effect of $\kappa$ on $\Delta V_{out}$ is more drastic when $\kappa < 10$ for $f = 1M\Omega$, while for $f = 100K\Omega$, effect on $\Delta V_{out}$ for $\kappa > 10$ is more drastic. For smaller $f$ ($10K\Omega$), variation of $\kappa$ has a negligible effect on $\Delta V_{out}$.

### 3.3. Trends with Channel Dimensions:

The analysis shown so far has assumed an ideal case where the cell volume is identical to the detection volume. This is although desirable but in practice may not be possible to implement due to variable cell size distribution of the target cells. Though the qualitative trends for $\Delta V_{out}$ are expected to be the same, the structural non-idealities could affect the cytometer's output signal. These structural non-idealities are incorporated as resistances in parallel ($R_p$) and series ($R_s$) as shown in Fig. $1d$. For calculating $V_{out1}$, we have assumed that the cell is positioned exactly in the middle of the detection volume.

In Fig. $6a$, $|\Delta V_{out}|$ for a target cell of $15\mu m$ diameter are shown as the detection volume is varied from $15\times 15\times 15\ \mu m^3$ (ideal case) to $50\times 50\times 50\ \mu m^3$. The maximum $|\Delta V_{out}|$ is obtained for the ideal case while a decrease in $|\Delta V_{out}|$ is observed as detection volume is increased. This is expected because a non-ideal detection volume adds additional current paths between the microelectrodes other than that through the cell thereby decreasing the electric field intensity across the cell and hence the $\Delta V_{out}$.

### 3.4. Effect of Cell parameters:

The analysis done so far has been done without varying the cell impedances shown in the cell model (Fig. 1). In practice, however, the cell impedances may vary across a wide range based on the size, type, and, morphology of the cell. For example, for complete blood counting ($CBC$) test, a microfluidic cytometer needs to simultaneously enumerate Red Blood Cells ($RBCs$), White Blood Cells ($WBCs$), and, platelets, etc. [15]. Moreover, the sub-populations of lymphocyte, granulocytes, and, monocytes in $WBC$ can be separated out from the electrical characteristics. In such experiments, $\Delta V_{out}$ and electrical opacity from microfluidic cytometers have often been used to distinguish between various types of blood cells as well as the sub-populations of the cells. In this section, we explore $\Delta V_{out}$ and opacity characteristics as a function of varying the morphology and size of the target cells.

For a given size of the cell, the morphological variations may change $C_m$ and $C_n$ without substantially affecting the cell resistances. Fig. 7 show the $\Delta V_{out}$ characteristics for varying the cell capacitances for two different microelectrode areas ($A_1 = 15 \times 15 \mu m^2$ and $A_2 = 200 \times 200 \mu m^2$) keeping all other cell/device parameters constant. For $A_1$, the effect of $C_m$ is significant in range of $f$ between $0.1 MHz - 4 MHz$. For $A_2$, however, the effect of varying $C_m$ is significant in a range of $f = 2.5 KHz - 4 MHz$. Fig. 7 $e - f$ shows the $\Delta V_{out}$ characteristics for $WBC$ (15$\mu m$), $RBC$ (6$\mu m$), and, platelets (3 $\mu m$) keeping $A = 15 \times 15 \mu m^2$. Here, since the cell size is varying, cytoplasm and nuclear resistances are also changed accordingly along with the cell capacitances. Due to change in cell resistances, the high frequency ($f > 1 MHz$) response for $|\Delta V_{out}|$ also gets modified as observed in the Fig. 7 $e$.

Fig.8 shows modulation in cell opacity ($OP$) for cell capacitance variations (Fig. 8 $a - f$) and for varying cell size (Fig. 8 $g - i$). Two different approaches to define $OP$ are compared in the three columns of Fig. 8. $OP1$ is defined as $\frac{V_{out1}(LF)}{V_{out1}(HF)}$ while $OP2$ is defined as $\frac{\Delta V_{out}(LF)}{\Delta V_{out}(HF)}$) where $HF$ is the high frequency which is kept constant at $1 MHz$ and $LF$ is the low frequency that is varied in Fig. 8. For $OP1$, that is shown in the left most column (Fig. 8 $a, d, g$) the effect of $C_{dl}$ is not substantial which results in $OP1$ being independent of the microelectrode area. For $OP2$, on the other hand, which shown in the middle column (Fig. 8 $b, c, e$) for $A = 15 \times 15 \mu m^2$, and in the right most column (Fig. 8 $f, h, i$) for $A_2 = 200 \times 200 \mu m^2$, the effect of $C_{dl}$ is significant and therefore a prominent effect of the microelectrode area can be observed. Since opacity is often used to distinguish cell populations on the basis of membrane properties (irrespective of cell size), the effect of microelectrode dimensions should be carefully considered when the differential voltage (impedance) signal from the cytometer readout circuit are used to calculate opacity. The most significant effect on $OP$ is observed for varying $C_m$ (Fig. 8 $a - c$) while the variation in $C_n$ shows a minor effect on $OP$.

## 5. Conclusions:

The physical properties of the target biomarker and the structural design of the microfluidic counter have significant effect on the output response of the coulter microfluidic cytometer. We have used circuit simulations based on equivalent cell model to quantify the impact of various design parameters for the device and the target cell characteristics on the output response of the device. The physical dimensions of the microelectrodes can strongly influence the output response of the sensor as a function of the frequency of the applied electrical stimulus. In particular, the lower frequency response can be significantly modulated with increasing microelectrode surface area due to the effect of double layer capacitance. This trend is shown to saturate at microelectrode area close to $200 \times 200 \mu m^2$. The detection volume should ideally be close to the size of the target cell to maximize the sensitivity. A detection volume bigger than the target cell size lowers the amplitude of higher frequency response while lower frequency response is less affected. The dielectric properties of the suspension medium and its electrical impedance can affect the lower and higher frequency response of the counter respectively. The resistance of the readout bridge circuit should be optimally selected based on the choice of the operating frequency. The cell's electrical opacity responds to the modulation in membrane capacitance for a specific range of signal frequencies that can strongly vary as a function of the microelectrode area and the choice of absolute cell impedance vs. the differential output impedance for opacity calculation. With optimally designed device structure, a microfluidic counter can effectively sense the target cells at signal frequency as low as few $KHz$. These insights can provide valuable guidelines for the design and characterization of coulter based microfluidic sensors.

## 6. Funding Information

This study was funded by Technology Development Fund (TDF02-127) from Higher Education Commission of Pakistan and Faculty Initiative Fund (FIF) from Lahore University of Management Sciences, LUMS.

## 7. Conflict of Interest

Authors of this study have no conflicts of interest to declare.

## 8. Supplementary Material

LTSPice netlist used in the simulations is shown below:

```
"ExpressPCB Netlist"
"LTspice XVII"
1
0
0
""
""
""
"Part IDs Table"
"Rn" "133.33K" ""
"Rc2" "133.33K" ""
"Cn1" "2.01e-12" ""
"Cn2" "2.01e-12" ""
"Rc1" "16.67K" ""
"Rc3" "16.67K" ""
"Cm2" "4.02e-12" ""
"Cm1" "4.02e-12" ""
"V1" "SINE(0)" ""
"C2" "9e-11" ""
"C1" "9e-11" ""
"Rout1" "10K" ""
"Rsol" "41666.67" ""
"C3" "9e-11" ""
"Rout2" "10K" ""

"Net Names Table"
"N002" 1
"N001" 3
"N005" 5
"N004" 8
"N003" 11
"N006" 13
"N007" 15
"N009" 18
"N012" 20
"0" 22
"N010" 25
"N008" 27
"N011" 29

"Net Connections Table"
1 1 1 2
1 4 2 0
2 1 2 4
2 3 1 0
3 2 1 6
3 4 1 7
3 6 2 0
4 2 2 9
4 3 2 10
4 5 1 0
5 5 2 12
5 8 1 0
6 6 1 14
6 7 2 0
7 7 1 16
7 10 1 17
7 13 2 0
```

```
8  8  2 19
8  11 1 0
9  9  1 21
9  10 2 0
10 9  2 23
10 12 2 24
10 15 2 0
11 11 2 26
11 12 1 0
12 13 1 28
12 14 1 0
13 14 2 30
13 15 1 0
```

11

**Figure Captions:**

**Fig. 1 a** Illustration of a biological cell **b** Electrical circuit model for the cell **c** Illustration of cell flow through microelectrodes in the detection volume **d** Overall electrical circuit for the cell counter

**Fig. 2 a** Cell impedance (amplitude and phase) vs. frequency **b** $V_{out1}$ (amplitude and phase) vs. frequency, **c** $V_{out2}$ (amplitude and phase) vs. frequency **d** $\Delta V_{out}$ (amplitude and phase) vs. frequency

**Fig. 3 a** $|\Delta V_{out}|$ vs. frequency for varying microelectrode area **b** $\angle \Delta V_{out}$ vs. frequency for varying microelectrode area **c** $|\Delta V_{out}|$ vs. microelectrode area at two different frequencies **d** $|\Delta V_{out}|$ vs. frequency for varying $R_{out}$ **e** $\angle \Delta V_{out}$ vs. frequency for varying $R_{out}$ **f** $|\Delta V_{out}|$ vs. $R_{out}$ for different signal frequencies

**Fig. 4 a** $|\Delta V_{out}|$ vs. frequency for different suspension medium **b** $\angle V_{out}$ vs. frequency for different suspension medium

**Fig. 5 a** $|\Delta V_{out}|$ vs. frequency for varying $\kappa$ **b** $\angle \Delta V_{out}$ vs. frequency for varying $\kappa$ **c** $|\Delta V_{out}|$ vs. $\kappa$ at different signal frequencies

**Fig. 6 a** $|\Delta V_{out}|$ vs. frequency for varying detection volume **b** $\angle \Delta V_{out}$ vs. frequency for varying detection volume **c** $|\Delta V_{out}|$ as a function of detection volume for variable frequencies

**Fig. 7 a** $|\Delta V_{out}|$ for varying cell membrane, $C_m$ **b** $\angle \Delta V_{out}$ for varying cell membrane, $C_m$ **c** $|\Delta V_{out}|$ for varying nuclear membrane, $C_n$ **d** $\angle \Delta V_{out}$ for varying nuclear membrane, $C_n$ **e** $|\Delta V_{out}|$ for different cell types **f** $\angle \Delta V_{out}$ for different cell types

**Fig. 8** Opacity and ΔOpacity as a function of the low frequency for: **a, b, c** Varying $C_m$, **d, e, f** Varying $C_n$ **g, h, i** Varying cell types. For the left column (**a, d, h**), opacity is evaluated using the ratio of the absolute change in the cell impedance at low vs. high frequency, while the ratio of differential impedance at low vs. high frequency is used for middle (**b, e, h**) and right (**c, f, i**) columns. The middle and right columns use different area for the microelectrodes